\documentstyle[12pt,aasms4]{article}

\begin{document}
\oddsidemargin=15mm
\onecolumn 

\title{Monte Carlo Simulations of Type Ia Supernova Observations in 
Supernova Surveys}

\author{Weidong Li,  Alexei V. Filippenko}
 \affil{Department of Astronomy, University of California, Berkeley, CA 94720-3411}

\bigskip
\centerline{and}

\author{Adam G. Riess}
\affil{Space Telescope Science Institute, 3700 San Martin Drive, Baltimore, MD 21218}

\newpage

\begin{abstract}

We have performed Monte Carlo simulations of type Ia supernova (SN Ia)
surveys to quantify their efficiency in discovering peculiar overluminous
and underluminous SNe Ia. We determined how the type of survey
(magnitude-limited, distance-limited, or a hybrid) and its characteristics
(observation frequency and detection limit) affect the 
discovery of peculiar SNe Ia. We find that there are strong biases against
the discovery of peculiar SNe Ia introduced by at least four observational
effects: the Malmquist bias, the age of the SN Ia at the time of 
its discovery, the shape of its light curve, and its degree of extinction.
Surveys with low observation frequency (less than once per 10 days) tend to 
discover SNe Ia which are too old for observers to easily recognize their peculiarity.
Subluminous SNe Ia are underrepresented in magnitude-limited surveys 
because they can only be found within a small volume and they remain
above the detection limit for less time. Conversely, overluminous SNe Ia
are more easily found in magnitude-limited surveys, although their likely 
association with dusty regions reduces the volume in which they can be
discovered. The unbiased rate of peculiar SNe Ia can be recovered only
in distance-limited surveys with high observation frequencies and with 
detection limits which are fainter than the peak magnitude of a subluminous
SN Ia in the farthest potential host.

\end{abstract}


\section{Introduction}

Since supernovae were first recognized by Baade \& Zwicky (1934) as a class of objects distinct
from common novae, over 1,600 of them have
been discovered by professional and amateur astronomers.
A record high of over 200 SNe were discovered in 1999.

Most of the SNe were found during systematic SN searches
(also known as SN surveys), the first of which was conducted by 
Zwicky using the 0.46-m Schmidt telescope at Palomar Observatory
(the Palomar SN Survey; Zwicky 1938; 
Kowal et al. 1974). There have been a number of successful SN 
surveys in the past, such as the Asiago SN survey (Ciatti \& Rosino 1978),
the Berkeley Automated SN Search (BASS; Perlmutter et al. 1992), and the Cal\'an/Tololo SN survey (CTSS; Hamuy et al. 1993).
Many SNe were also discovered in the course of the first generation
and second generation Palomar Observatory Sky Surveys (POSS I \& 
II; e.g., Reid et al. 1991) and the UK Schmidt Sky Survey (e.g.,
Corwin, de Vaucouleurs, \& de Vaucouleurs 1977). The number of active
professional SN surveys is currently at an all-time high:
the Lick Observatory SN Search
(LOSS, Treffers et al. 1997; Li et al. 2000a; Filippenko et al. 2000), the Beijing Astronomical Observatory SN Survey
(BAOSS; Li et al. 1996),
the Perth Observatory SN Search (Williams et al. 1995),
the Nearby Galaxies SN Search (Strolger et al. 1999), 
the High-Z SN Search (HZSS; e.g., Schmidt et al. 1998), the SN
Cosmology Project (SCP; e.g., Perlmutter et al. 1997),
the Mount Stromlo Abell Cluster SN Search (MSACSS;
Reiss et al. 1998), the Wise Observatory Optical Transients Search (WOOTS; Gal-Yam \& Maoz 1998),
the EROS experiment (e.g., Palanque-Delabrouille et al. 1998), and others.
Amateur astronomers, armed
with modern, inexpensive CCD cameras and moderately large telescopes, 
have contributed significantly to the discoveries of SNe.  To name a 
few, the UK nova/SN patrol (e.g., Hurst \& Armstrong 1998), the Tenagra Observatories
(Schwartz 1997), and the Puckett Observatory (Puckett 1998) are all quite successful in discovering
SNe. 

Most of these SN surveys can be described using two categories based on how the 
survey fields are selected: magnitude-limited and distance-limited.
In a magnitude-limited SN survey, the target fields are usually random regions
on the sky that contain many galaxies at different redshifts. The number 
of SNe discovered depends on the detection limit of the images 
(hence magnitude-limited). The CTSS, HZSS, 
SCP, and 
EROS experiment are examples of magnitude-limited SN surveys.
In a distance-limited SN survey, in contrast, the target fields are usually
individual 
galaxies or clusters of galaxies. The limiting magnitude of the survey images may be much deeper than
all the possible SNe in the sample galaxies at peak (as long as they are not heavily 
extinguished), so the number of SNe discovered largely depends on the 
number and distances of the sample galaxies or clusters (hence distance-limited).
LOSS, BAOSS, and BASS are examples of distance-limited SN surveys. A few
SN surveys, however, are best described as a hybrid of these
two categories. For example, MSACSS searches a number of Abell 
clusters of galaxies with fixed redshifts, 
but half of the SNe were actually discovered in the background of the
clusters and some of them have magnitudes 
close to the detection limit of the survey images. The SNe discovered
in the target galaxy clusters are distance-limited but the background
ones are magnitude-limited.
 
The discovery and observation of Type Ia SNe (SNe Ia), generally
the most luminous and homogeneous 
SNe, have been increasingly 
emphasized in modern SN surveys because of their utility
for cosmological studies (see Riess et al. 1999a, and references therein).
In particular, HZSS, SCP, MSACSS, and WOOTS 
are all designed to search for SNe Ia. SNe Ia have been used to 
measure the current expansion rate of the Universe, study the 
peculiar velocities of distant galaxies as well as the bulk flow of our 
own local neighborhood, examine the nature of the redshift using the
time dilation test, probe the nature of extragalactic dust, study
the chemical evolution of galaxies, and measure the expansion history 
of the Universe.

Although SNe Ia are more homogeneous than other classes of SNe,
diversity among them was recognized and has been quantified
during the last decade. {It is now generally thought that 
SNe Ia can be subclassified by their spectral characteristics as SN 1991T-like,
normal, and SN 1991bg/1986G-like (hereafter SN 1991bg-like) objects (see 
Filippenko 1997 for a review, and references therein; Leibundgut 2000). 
Both the SN 1991T-like and the SN 1991bg-like objects
are also called ``spectroscopically peculiar SNe Ia." SN 1991T-like
objects are peculiar as their spectra do not show the classic
Si II 6150 \AA\, line of normal SNe Ia but show strong Fe III
lines. SN 1991bg-like objects are peculiar as their spectra
show an enhanced Si II 5800 \AA\, absorption and a broad absorption
trough extending from about 4100 to 4400 \AA\,(probably due to Ti II
lines). Li et al. (2000c) point out the existence of a class of ``SN 1999aa-like" objects,
which are spectroscopically similar to SN 1991T-like objects but also show apparent
Ca II H\&K lines that are weak in SN 1991T. Whether those SN 1999aa-like
objects are the missing link between SN 1991T-like events and normal SNe
Ia remains to be determined. In this paper we do not differentiate between
SN 1991T-like and SN 1999aa-like objects, instead calling all of them 
SN 1991T-like.}

The variations among light-curve shapes, luminosity,  and spectral evolution 
of the different major subtypes of SNe Ia introduce several observational
biases in SN surveys, which can profoundly affect the observed
peculiarity rate and the observed luminosity function of SNe Ia
(the number distribution of SNe Ia among the three subtypes). 
Since the intrinsic peculiarity rate and the true
luminosity function of SNe Ia hold important clues to their 
theoretical models, and since it is important to understand
the completeness of different kinds of 
SN Ia surveys, we have written Monte Carlo codes to study the 
discovery {efficiency} of SNe Ia in both the distance-limited and the 
magnitude-limited SN surveys. The preliminary results of the simulations
were reported by Li et al. (2000b), but the final results are 
described here. These will also be used in a companion paper (Li et al. 2000c),
where the intrinsic peculiarity rate and the intrinsic luminosity function of 
SNe Ia are investigated with the sample of SNe Ia discovered during
the course of LOSS and BAOSS.  Hamuy \& Pinto (1999) 
have used a similar approach to study the observational 
biases in CTSS,  and we  
compare our results with theirs. 

This paper is organized as follows. Sec. 2 presents the 
various observational biases for SNe Ia in SN surveys, 
and Sec. 3 reports the outcomes of the Monte Carlo simulations.
We discuss our results in Sec. 4 and summarize our conclusions in Sec. 5.

\section{ The Observational Biases}

The differences among the luminosities, light-curve
shapes, and spectroscopic evolution of the three major subtypes of
SNe Ia result in four main observational biases in SN 
surveys:  the ``age bias,"
the Malmquist bias, the ``light-curve shape (LCS) bias," and the 
``extinction bias."

\subsection{The Age Bias}

The spectroscopic peculiarity of SN 1991T-like objects can be easily detected only prior to
or near maximum brightness (e.g., Filippenko et al. 1992a; Phillips et al. 1992); thus, we cannot be sure whether the SNe discovered more
than a week after maximum are peculiar (SN 1991T-like). If these 
SNe are counted as normal in the sample, it will underestimate the 
peculiarity rate and change the luminosity function of SNe Ia.

\subsection{The Malmquist Bias}

SN 1991T-like objects are intrinsically brighter by $\sim$0.4 mag at peak (e.g., Filippenko et al. 1992a; Phillips et al. 1992) 
and SN 1991bg-like objects are intrinsically fainter by $\sim$2 mag (e.g., Filippenko et al. 1992b; Leibundgut et al. 1993) than normal SNe Ia
in the $B$ band. The difference in other photometric
bands (e.g., $V$ or $R$) varies and is discussed in more detail
in the next section.
This difference in the luminosity results in a Malmquist bias that
affects the observed peculiarity rate and luminosity function of SNe Ia.

Assume all three kinds of objects suffer equal amounts of extinction. 
A magnitude-limited supernova survey will then overestimate
the rate of SN 1991T-like objects because of the larger volume
of space surveyed for such objects. 
It will also underestimate the
rate of SN 1991bg-like objects because of the smaller volume of
space surveyed for those objects.

The case for a distance-limited supernova survey is more complicated. 
Whether the Malmquist bias 
underestimates or overestimates the rates depends sensitively
on the characteristics of the survey such as the limiting distance, the 
limiting magnitude, and the baseline (the time interval
to repeat the observations).
More details about the Malmquist bias are addressed in the sections
on Monte Carlo simulations.

\subsection{ The Light-Curve Shape Bias}  

There is an observational bias caused by the differences among the light-curve
shapes of SNe. SN 1991T-like objects rise to their maxima and
decline thereafter more slowly than do normal SNe Ia; they have
the so-called ``slow light curves."  SN 1991bg-like objects rise to their maxima 
and decline thereafter more quickly than do normal SNe Ia; they
have the so-called ``fast light curves." The LCS bias affects the amount
of time a SN Ia is above the detection limit and its changing brightness
from one epoch to another, and hence the discovery efficiency and the distribution
of the epoch of discovery as well. These factors result in an observational
bias that affects the observed peculiarity rate and luminosity function
of SNe Ia.

\subsection{The Extinction Bias}

There is evidence that SN 1991T-like objects may suffer more extinction
than the other two kinds of objects because they are usually (perhaps always) seen in 
dusty, star-forming regions. For example, four known SN 1991T-like objects
all show a significant amount of reddening $E(B-V)$: SN 1991T has 0.13
mag (Filippenko et al. 1992a), SN 1995ac has 0.17 (Riess et al. 1999a),
SN 1995bd has 0.5 (Riess et al. 1999a), and SN 1997br has 0.35
(Li et al. 1999). When the Galactic component of the extinction is 
subtracted according to the map of Schlegel, Finkbeiner, 
\& Davis (1998), the intrinsic reddening due to the host galaxy of 
the SN is $E(B-V) =0.11$ mag for SN 1991T, 0.13 mag for 
SN 1995ac, 0.00 mag for SN 1995bd, and 0.24 mag for SN 1997br.
The mean reddening of these four SN 1991T-like objects is
$E(B-V) = 0.12$ mag, 
{higher than
the mean host-galaxy reddening of $E(B-V)$ = 0.05 mag
for 26 normal SNe Ia studied by
Phillips et al. (1999).} If the three major subtypes of SNe Ia indeed 
suffer different amounts of extinction, the observed peak apparent magnitude
will be affected and an observational bias similar to the Malmquist bias 
results.

\section{ The Monte Carlo Simulations}

\subsection{Assumptions}

To study the roles of various observational biases, we have written
a Monte Carlo code to simulate magnitude-limited and distance-limited
SN surveys.  We make a number of assumptions in our study:

(1) SNe Ia are distributed uniformly in space. 

(2) The absolute magnitude for a normal SN Ia is $-$19.5 (in the $R$ band; 
see discussions below).

(3) We assume that the objects have the $R$-band light curves shown in Figure 1. 
We choose to use the $R$ band
because most modern CCD SN surveys are done in unfiltered
mode, whose closest match to the standard passbands is $R$. 
We have 
constructed the $R$-band light curve of SN 1991T-like objects from 
the data of SN 1991T (Lira et al. 1998), SN 1995ac (Riess et al. 1999a),
and SN 1998es (Li et al. 2000d). A 6-order cubic spline is used to 
fit the data of the 3 SNe from $-$13 days to 80 days. We have used 
the data of SNe 1998de, 1999by, and 1999da (Li et al. 2000e) to 
construct the $R$-band light curve of SN 1991bg-like objects
from $-$11 to 80 days. The $R$-band light curve
of SN 1994ae from $-$12 to 80 days (Riess et al. 1999a) is adopted
as that of the normal objects. The rising parts of the light curves of
these SNe are constructed according to the recipe described by Riess et al.
(1999b), where we adopted risetimes of 24.0, 21.0, and 18.0 days for 
SN 1991T-like, normal, and SN 1991bg-like objects, respectively.
Those risetimes are their counterparts in the $B$ band (Riess et al.
1999b) plus 2.0, 2.0, and 1.5 days for the SN 1991T-like, normal, and 1991bg-like
objects, respectively; observations usually show that the $R$-band
maximum follows the $B$-band maximum by about 2 days. As the adopted
light curves of the SNe play an important role 
in our simulations, in Sec. 4 we will investigate the results of using 
light curves in other passbands.  Following convention, the time of 
$B$ maximum brightness is defined as t = 0 (or epoch = 0) throughout 
this paper. 

(4) There is 0.35 mag of scatter in the absolute magnitudes 
of each type of SN Ia in the $R$ band. {This scatter is chosen to be similar to that found
in the $B$ and $V$ bands for normal SNe Ia (e.g., Hamuy et al. 1996c).}
In the simulations,
the scatter is introduced by changing the magnitude of each SN by a
normally distributed number with 1$\sigma$ = 0.35 mag. {Note 
that this simple luminosity function for normal SNe Ia (a Gaussian
distribution with 1$\sigma$ = 0.35 mag, centered at $-$19.5 mag) 
is different from that found by Hamuy \& Pinto (1999) for 
CTSS, which is quite flat from $-$19.0 to 
$-$19.9. However,
since there is no luminosity function available for the SN 1991T-like
and SN 1991bg-like objects, we choose to use a Gaussian
distribution for all three kinds of objects to be consistent.}

(5) We assume that SN 1991T-like objects are intrinsically 0.4 mag brighter, 
and SN 1991bg-like objects are intrinsically 1.0 mag fainter, than normal SNe Ia in the $R$ band.
The peak apparent magnitudes in the $B$ and $R$ bands after correction 
of extinction are about the same for the SN 1991T-like objects (Lira et al. 1998; Riess et al. 1999a; Li et al. 2000d) and the normal objects (Riess et al. 1999a), so we have adopted their difference of luminosity in the $B$ band (0.4 mag)
to be that in the $R$ band. The peak apparent magnitude of SN 1991bg-like objects
in the $R$ bands is found to be about 1 mag brighter than that in the $B$
band (Li et al. 2000e), so we derive that they are about 1.0 mag fainter
than the normal SNe Ia in the $R$ band if we assume that there is a 
2.0 mag difference in the $B$ band. 

(6) 
To study the effects of extra extinction for SN 1991T-like
objects (the extinction bias), we have modeled three cases:
(a) they have $R$-band extinction 
$A_R = 2.5 E(B-V)$
equal to that of the other two kinds of objects [hereafter the
case SN 1991T ($A$=0)]; (b)
they suffer 0.4 mag more extinction than the other two kinds of objects [hereafter the
case SN 1991T ($A$=0.4)];
and (c) they suffer 0.8 mag more extinction than the other two kinds of objects
[hereafter the case SN 1991T ($A$=0.8)].

(7) We assume that a SN is detected when its magnitude is brighter than 
the limiting magnitude of the survey {by a 
small number drawn from the positive part of a 
normal distribution with 1$\sigma = $  0.3 mag}. 
There are several reasons why the SN has to be {\it brighter 
than } the limiting magnitude to be discovered: (a) there is extinction
in the Galaxy and in the host galaxy of the SN; (b) the weather may not always
be good enough for the surveys to reach their limiting magnitudes; (c) SNe
often occur in complex regions of the host galaxy, making them difficult to
discover at the limiting magnitude. {Note that the situation of 
real observations is often more complicated than our simple model of 
accounting for extinction, weather, and background. In particular, 
the extinction to the SN depends on both the Galactic extinction
to its host galaxy (which is constant) and the supernova's  relative
position within the host (close to spiral arms, etc.), and weather
conditions usually have their own patterns. However, since our simulations
are not done for a particular survey
with known weather patterns, there is no readily available mathematical
description of the real situation and we have chosen to do the most
general case.}

(8) We assume that when a SN 1991T-like object is discovered more than 7 days 
past maximum brightness, it is not distinguished as a SN 1991T-like
object, but rather
is classified as a normal SN Ia by the observer because the spectrum
looks normal. The cutoff of 7 days past maximum brightness is arbitrary,
and is used to demonstrate the effect of the age bias. In the 
discussion section we will explore the results of the simulations using 
different cutoffs.

(9) We  do not simulate the position of the SN in the host galaxy,
which means that the discovered SN does not need to have a minimum
projected separation from the galactic nucleus {or other regions
of high surface brightness like H II regions or spiral arms}. 
This is generally acceptable for CCD SN surveys, except perhaps in the 
central few arcseconds of galaxies with starlike nuclei, 
but unacceptable for photographic surveys as discussed by Hamuy \& Pinto (1999).
We also do not simulate the Hubble types of the host galaxies of the SNe.

(10) We do not take into account practical limitations on 
telescope time. A supernova survey
with a limiting magnitude of 19.0 and a baseline of only one day will need
more than the available time per night when there are many fields in the sample.
In our simulations, we assume all  galaxies can indeed be observed with the
desired frequency.

(11) We do not take into account the effect of the seasons. In other words,
all the SNe used in the simulations are those {\it observable} by the observers.
In practice, most galaxies will have a long period of time (about 5 months)
when they not observable because of their proximity to the Sun. Since
the effect of the seasons is similar to decreasing the input number of 
all three kinds of SNe proportionally, we do not consider it in our simulations.

{(12) We do not consider the effect of the redshift (the ``$K$-correction"),
the time-dilation effect, or the geometry of space.
All these effects are negligible in our simulations of nearby SNe Ia.
}

{(13) We do not apply the luminosity vs. light-curve shape correlation (e.g.,  
Phillips et al. 1999; Riess et al. 1998; Perlmutter et al. 1997), beyond
simply adopting different average light curves for the three subclasses
of SNe Ia. Our simulaitons are done in the $R$ band, where there is 
little informaiton  for variations within a given subclass.}

{(14) Generally there is a gap (2 to 5 days, sometimes longer) 
between the time of a SN discovery and the time of its spectroscopic
classification. Since the peculiarities considered here are spectral, 
it is important to simulate this fact
to study the age bias. In the simulations the time of the spectroscopic
classification of a SN is taken to be its discovery
epoch plus an offset of a random time between 2 and 5 days.} 

(15) For convenience, we need to choose an initial luminosity function of
SNe Ia. In our simulations,
we assume that SN 1991bg-like and SN 1991T-like objects have the same number
density, and that normal SNe Ia have a number density that is four times higher than 
either of them. {This choice of luminosity function is quite arbitrary  and
is used only to demonstrate the effects of the various observational biases
on the detection efficiency of different kinds of objects (nevertheless,
a luminosity function similar to this is found by Li et al. [2000c])}. {Choosing 
a different luminosity function will not change the major results of the 
simulations. In particular,
we will evaluate the effects of the observational bias on the 
SN 1991bg-like objects by considering SN 1991bg-like and normal 
objects only (which means an input rate of SN 1991bg-like objects of 20\%);
the same is true for dealing with SN 1991T-like objects.
The intrinsic luminosity function of SNe Ia is discussed
in the companion paper by Li et al. (2000c).}

\subsubsection{Magnitude-Limited Supernova Survey}

{Throughout this paper distances are expressed in terms of the 
distance modulus ($\mu_0$).
The apparent magnitude ($m$) of a normal
SN Ia at a distance modulus $\mu_0$ is $m_{normal} = \mu_0 - 19.5$.

The limiting magnitude of the survey is set to be 19.0. This is an
arbitrary choice and the results can be scaled. 
We use 20,000,000 normal SNe Ia uniformly distributed in space to
a distance of $\mu_0$ = 39.55 mag, where a normal SN 
Ia has an apparent magnitude
of 20.05. Because of the scatter of magnitudes, we need to consider
distances where normal SNe are fainter than the limiting magnitude.
We adopt $\mu_0$ = 39.55 mag because the resulting magnitude for a normal 
SN Ia (20.05) is the limiting magnitude of 19.0 plus
3$\sigma$ (1.05 mag) of the scatter in magnitudes. }

Since SN 1991T-like objects are more luminous by 0.4 mag than 
normal SNe Ia, the survey can detect SN 1991T-like objects to 
 $\mu_0$ = 39.95 mag (where normal SNe Ia reach 20.45 mag) for the case SN 1991T ($A$=0). 
Thus, the space volume in which SN 1991T-like objects ($A$=0)
can be detected is 1.738 times that of the normal SNe Ia. With our
adopted number density this means that there are 8,690,000 SN 1991T-like
objects observable by the survey.

For the case SN 1991T ($A$=0.4), SN 1991T-like objects have 
apparent magnitudes that are the same as those of the normal 
objects. Hence, there are 5,000,000 SN 1991T-like 
objects observable by the survey.
For the case SN 1991T ($A$=0.8), the survey can only detect SN 1991T-like objects to 
a distance of $\mu_0$ = 39.15 mag (where normal SNe Ia reach 19.65 mag). The space volume in which
SN 1991T-like objects can be detected is then only 0.575 times that of 
the normal SNe Ia, which means that there are 2,875,000 SN 1991T-like 
objects observable by the survey.

Likewise, the space volume in which SN 1991bg-like objects can be detected
is only 0.251 times that of the normal SNe Ia, which means that there 
are 1,255,000 SN 1991bg-like objects observable by the survey.

All of these SNe are binned into 0.01 mag intervals, and their number 
distributions are shown in Figure 2. This is the input to the
Monte Carlo simulation of the magnitude-limited
supernova survey.

Even at the time of the input, the Malmquist bias dramatically changes
the observed relative frequencies of peculiarity in some cases.
The percentage of peculiar SNe Ia relative to normal objects becomes 5.0\% for SN 1991bg-like
objects (20\% intrinsic); and 30.3\% for SN 1991T ($A$=0), 20.0\% for 
 SN 1991T ($A$=0.4), and 12.6\% for SN 1991T ($A$=0.8) (20\% 
intrinsic).

{The only parameter in the simulation of the magnitude-limited survey
is the baseline. We have simulated results for baselines $t_b$ = 1, 3, 
5, 7, 10, 15, 20, 25, 30, 40, 50, and 60 days.
For a given baseline ($t_b$), a SN in a magnitude bin is given a 
normally distributed scatter with $1\sigma= 0.35$ mag, and 
the limiting magnitude of the survey is modified (made brighter) by a
number drawn from the positive part of a 
normal distribution with $1\sigma = 0.3$ mag. The SN is then assigned a random 
observation time ($t_{obs}$) between 0.0 and $t_b$ before maximum.
With $t_b$ and $t_{obs}$ the simulation goes through the light curve of the
SN and finds the brightest point at which the 
SN would be observed in the series of images $t_{obs}$, $t_{obs}+t_b$, $t_{obs}+2*t_b$, ..., 
to see whether it is brighter than
the modified limiting magnitude; if it is, the SN is marked as being detected. 
The simulation also goes through the light curve of the SN to find the earliest time that
the SN could have been discovered (e.g., time series of $t_{obs}$, $t_{obs}-t_b$,
$t_{obs}-2*t_b$, ...); this is 
the epoch of discovery for the SN.} 

The results of our simulation are shown in Figures 3 through 6. Figure
3 illustrates the number of SNe discovered at different apparent magnitudes.
{Notice that the dotted lines marked ``perfect" in the figure represent the 
number of SNe versus apparent magnitude at maximum brightness {\it after}
adding Gaussian scatter with $1\sigma = 0.35 $ mag, not the input
distributions without adding scatter as in Figure 2.}
Here we emphasize again that each subclass of peculiar SNe Ia is studied
by comparing that subclass with the normal subclass only. Figure 3 contains two 
of the four cases: SN 1991bg-like objects and the case SN 1991T ($A$=0).
The upper panel shows the number distribution of the peculiar SNe Ia,
while the lower panel shows the corresponding number distribution of the
normal SNe. 
The other two cases [SN 1991T ($A$=0.4) and SN 1991T ($A$=0.8)]
are not shown, but their results are similar to the case SN 1991T ($A$=0).

We notice the following. (1) Even with a baseline of only one day, not all SNe can be 
discovered, and the fainter the apparent magnitude, the more the SNe are 
lost in the survey. The discovery efficiency (the number of SNe
discovered divided by the input number of SNe with mag $\le$ 19.0; see Figure 4)  
can only reach a maximum of 84\% for all three kinds of objects.
(2) The longer the baseline, 
the more SNe are lost in the survey. (3) For the case SN 1991T ($A$=0),
when the baseline is 30 days, the 
number of normal SNe detected with apparent magnitude brighter than 
about 18 is slightly more than the number of normal SNe Ia at
that magnitude on the ``perfect" line. This is 
caused by the fact that with such long baselines, 
some SN 1991T-like objects are discovered later than one
week past maximum and are counted as normal SNe Ia due to the age bias.

Figure 4 shows the discovery efficiency at different baselines. The upper two panels 
display the results of the SN 1991bg-like and the SN 1991T-like objects, while 
the bottom panel shows the result of SN 1991T-like objects without considering
the age bias. Things to 
notice are as follows. (1) The discovery efficiency declines quickly with increasing
baselines. (2) In the middle panel, 
all three cases ($A$=0, $A$=0.4, and $A$=0.8) have the same discovery efficiency 
for the SN 1991T-like objects, but the corresponding normal objects have
different discovery efficiencies at long baselines. This is caused by 
the fact that when the extinction is
bigger, there are fewer SN 1991T-like objects 
observable by the survey; consequently, there are fewer
SN 1991T-like objects discovered later than 
a week past maximum and counted as normal.
The case SN 1991T ($A$=0.4) (the dashed line) thus
has a smaller discovery efficiency for the normal objects, 
while the case SN 1991T ($A$=0.8) (the dotted line) has the smallest.
(3) When the baseline is shorter than 10 days, 
the normal objects have a larger discovery efficiency than the SN 1991bg-like
objects, but the SN 1991T-like objects have a slightly larger discovery efficiency 
than the normal objects. 
The main reason for this is the different light-curve shapes. 
Among the three kinds of SNe Ia, the SN 1991T-like objects spend the longest time
bright because of their broad light curves, and thus have the largest
discovery efficiency. (4) When the baseline is longer than 10 days, however, 
the normal SNe Ia have a larger discovery efficiency than the SN 1991T-like 
objects. The main reason for this
is the fast decline of the discovery efficiency of the SN 1991T-like objects,
which results from the age bias;
some of the SN 1991T-like objects are discovered later than a week past maximum
and are counted as normal SNe Ia. If we do not consider the age
bias in the simulations as shown in the bottom panel, 
the SN 1991T-like objects have a larger discovery efficiency
at all baselines.

Figure 5 shows the distribution of the epoch of discovery in the simulations,
where the epoch of maximum brightness is day 0. 
The case for SN 1991bg-like objects and the case SN 1991T ($A$=0) are displayed.
Again, the corresponding distributions for the normal SNe are shown in the lower
panel.
We see the following. (1) The smaller the baseline, the earlier is the 
peak in the discovery epoch. 
The epoch of discovery peaks at about $-$6, $-$11, and $-$12 days
for the SN 1991bg-like, normal, and  
SN 1991T-like objects, respectively, when the 
baseline is 1 day, and this changes to $-$2, $-$4, $-$6 days when the baseline
is 10 days. (2) {The majority of SNe Ia can be detected before maximum
only when the baseline of the survey is smaller
than 10 days}. (3) {For the SN 1991T-like objects, the distribution 
has a gradual cutoff from +2 to +5 days when the baseline is longer than 
30 days because of the age bias.
Since the cutoff of the age bias is +7 days, and there is a 
random gap of 2 to 5 days between the epoch of discovery and the 
time of spectroscopic classification, the discovery epoch of SN 1991T-like
objects should have this gradual cutoff from +2 to +5 days, and no 
SN 1991T-like objects should be identified by the observer
when their discovery epoch is later than +5 days. 
(4) The distribution of normal SNe Ia has a gradual increase from
+2 to +5 days and forms a second peak for the same reason. }
(5) For a given baseline, the SN 1991bg-like objects
have the narrowest peak in the distribution, and the SN 1991T-like
objects have the
broadest peak. This is caused by the difference in the light-curve shapes. 
The SN 1991bg-like objects have the narrowest light curve, so the detected
events have the largest probability of being discovered near maximum
brightness,
while the SN 1991T-like objects have the broadest light curve, so the detected
events spread out in the epoch of discovery.

Figure 6 shows the observed peculiarity rates in our simulations. 
The input rate for each type of SN is shown as dash-dotted lines. 
The difference
between the output peculiarity rates and their input values also
reflects the relative change in the luminosity function.
We see the following.
(1) The magnitude-limited surveys significantly underestimate
the rate of SN 1991bg-like objects for all baselines.
The rate changes from 20.0\% (the input peculiarity rate)
to about 5.5\%. (2) {For the case SN 1991T ($A$=0), the surveys
overestimate the rate of SN 1991T-like objects when the baseline
is smaller than 35 days, but underestimate it with larger baselines.
The rate changes from 20.0\% to more than 30\% at small baselines. The 
rate of SN 1991T-like objects first becomes larger with increasing
baseline, because SN 1991T-like objects are easier to discover due to
their broader light curves; it subsequently declines quickly with
increasing baselines because of the age bias.}
(3) For the case SN 1991T ($A$=0.4), SN 1991T-like objects have the same
luminosities as the normal objects, yet their rate is slightly larger than 
the input value when the baseline is small because of the LCS bias,
and is smaller than the input value when the baseline is large because
of the age bias. (4) For the case SN 1991T ($A$=0.8), the surveys
underestimate the rate of SN 1991T-like objects at all baselines (smaller
than 14\%). The rate of SN 1991T-like objects is less than 10\% when
the baseline is longer than 30 days. 

\subsubsection{Distance-Limited Supernova Survey}

{We assume the limiting distance is where normal SNe Ia have an apparent
magnitude at maximum of 
16.0 ($\mu_0$ = 35.5, $\sim$ 130 Mpc), and there are 4,000,000 normal
SNe Ia in that volume (down
to mag 16.0). With our arbitrary luminosity function for SNe Ia, 
there are 1,000,000 SN 1991T-like objects in the sample with magnitude down
to 15.6, 16.0, or 16.4 for the case SN 1991T ($A$=0.0), ($A$=0.4), or ($A$=0.8),
respectively, and there are 1,000,000 SN 1991bg-like objects with magnitude
down to 17.0. All the SNe are binned into 0.01 mag 
intervals, and their number distributions are shown in Figure 7.}

There are two parameters for the distance-limited survey: the limiting 
magnitude of the survey and the baseline. We have simulated results for 
limiting magnitudes of 15.0, 15.5, 16.0, 16.5, 17.0, 17.5, 18.0, 18.5, and 19.0.
The choice of baseline is the same as for the magnitude-limited
SN survey: 1, 3, 5, 7, 10, 15, 20, 25, 30, 40, 50, and 60 days.
The method of the simulations is basically the same as we described in 
the magnitude-limited SN survey, the only difference being that the limiting
magnitude is now a changing parameter.

Some of the results of the simulations are displayed in Figures 8 through 11.
Figure 8 shows the number of SNe discovered at different apparent peak magnitudes
for surveys with different limiting magnitudes. For each limiting magnitude,
results of different baselines are displayed. The results of the case SN 1991T
($A$=0.8) are shown here and the other cases yield similar results.

 Inspection of the figure reveals the following.

(1) When the limiting magnitude  of the survey 
is small (bright), such as 15.0, the results of
the distance-limited survey are exactly the same as those of a magnitude-limited 
survey.  In fact, for the input parameters of our distance-limited survey, it
will behave as a magnitude-limited survey as long as the limiting magnitude of 
the survey is $\le$ 15.6, the faintest SN 1991T-like
objects in the sample for the case SN 1991T ($A$=0).

(2) When the limiting magnitude equals 18.0 or 19.0, 
the number distributions of the distance-limited survey 
become  different from those of the
magnitude-limited surveys.  With a small baseline (1 to 10
days), all the SN 1991T-like objects and the normal SNe Ia are discovered
in the survey. When the baseline is large (e.g., 30 or 60 days), however,
some of the SNe are still lost in the survey. 

Figure 9 shows the discovery efficiencies for all cases. In each case
results for different baselines and 
limiting magnitudes are given. Rates for peculiar SNe Ia are
in the upper panel, and the corresponding rates for normal SNe Ia
are in the lower panel.  We notice the following.

(1) The discovery efficiency generally increases with increasing limiting magnitude. For example,
when the baseline equals 20 days, the discovery efficiency for the SN 1991bg-like objects
is only about 15\% for a limiting magnitude of 16.0, but it becomes 
56\%, 97\%, and 100\% for limiting magnitudes of 17.0, 18.0, and 19.0, 
respectively. 

(2) For the SN 1991bg-like objects, the survey discovers all the SNe when 
the limiting magnitude is 19.0 and the baseline is smaller than 30 days.
The discovery efficiency then declines with increasing baselines. For the other
limiting magnitudes, the discovery efficiency is always smaller than 1 and declines
with increasing baselines.

(3) For the case SN 1991T ($A$=0),  the survey discovers all the SNe  
when the baseline is small enough and the limiting magnitude is fainter 
than all the SNe in the sample. For example, when the limiting magnitude is
17.5, all SNe are discovered if the baseline is shorter than 20 days.
When the baseline becomes longer than that, the efficiency of SN 1991T-like
objects declines rapidly, while the corresponding efficiency of normal SNe Ia
first rises to above 100\%.
This is caused by the age bias 
because some of the SN 1991T-like objects are classified as normal SNe Ia
with such long baselines.

(4) Extinction of SN 1991T-like objects seems to play a negligible 
role in determining the discovery efficiency when the survey has a 
deep limiting magnitude (e.g., 17.5 and 19.0 in Figure 9) and
a small baseline, because all SNe are discovered regardless of the adopted
value for the extinction of SN 1991T-like objects. It only
affects the survey with a limiting magnitude that is comparable 
to or brighter than the faintest SNe in the sample (e.g., 16.0 in 
Figure 9), or when the baseline is long (e.g., $>$ 30 days).

Figure 10 shows the distribution of the epoch of discovery
for the case SN 1991T ($A$=0.8). Other cases have similar 
results.  When the baseline is small, such as 5 days, all the distributions
exhibit a Gaussian profile except the case of the SN 1991T-like objects
with limiting magnitude of 19.0, and the peak shifts toward later epoch of discovery
with decreasing limiting magnitudes. 
 When the baseline gets longer, the 
distribution becomes wider, and begins to show a flat-topped feature.
{The gradual increase for the normal SNe Ia and cutoff for the
SN 1991T-like objects at an epoch of +2 to +5 days when the baseline 
is 30 or 60 days is caused by the age bias, as discussed
previously.}  The distance-limited surveys with a baseline of 60 days and limiting
magnitudes of 18.0 or 19.0 are the only ones that have significant number 
distributions (e.g., the lower left panel of Figure 10)
at a discovery epoch of 20 to 40 days in our simulations. 

Figure 11 shows the peculiarity rates from our simulations of the distance-limited
surveys. The difference between the output peculiarity rates and their input
values also reflects the relative change in the luminosity function. We find the following.

(1) When the limiting magnitude of the survey is bright, such as 15.0,
all cases show a rate distribution that is exactly the same as that of the
magnitude-limited survey case (Figure 6). This is because when the limiting 
magnitude is so bright, the survey is actually limited by its limiting
magnitude rather than by the distances of the sample galaxies, so the survey
behaves as a magnitude-limited one. This kind of survey seriously 
underestimates the rate of SN 1991bg-like objects, overestimates 
the rate of SN 1991T-like objects for the case SN 1991T ($A$=0), overestimates
or underestimates the rate of SN 1991T-like objects for the 
case SN 1991T ($A$=0.4) depending on the baseline of the survey, and
underestimates the rate of SN 1991T-like objects for the case 
SN 1991T ($A$=0.8).

(2) When the limiting magnitude of the survey is 17.0, things become
more complicated. For SN 1991bg-like objects, there are many 
SNe with magnitude fainter than 17.0, so the rate should be
the same as that of the magnitude-limited case. Inspection of Figure 11 reveals, 
however, that this is not the case. The rate for SN 1991bg-like objects 
is always higher than in the corresponding magnitude-limited case. The reason for this
is that we calculate the rate of SN 1991bg-like objects by comparing 
them with the normal SNe Ia, which are not magnitude-limited 
(recall that all normal SNe Ia are brighter than mag 16.0). 
The number of normal objects discovered 
will always be smaller than in the magnitude-limited case (which have
normal SNe Ia observable to mag 17.0), resulting in a higher rate
for SN 1991bg-like objects. {The cases of SN 1991T ($A$=0) and ($A$=0.8)
have similar rate distributions. The rate equals the input value
when the baseline is small, and declines with increasing baselines.
The extinction seems to only affect the cutoff of the baseline
where the observed rate begins to decline from the input value: 20 days for the $A$=0 case
and 
15 days for the $A$=0.4 case. The SN 1991T ($A$=0.8) case has a 
rate distribution similar to that of the other SN 1991T cases,
but its rate is always smaller
than the input one.}

(3) When the limiting magnitude of the survey reaches 19.0,
a magnitude that is fainter than all the possible SNe in the 
sample galaxies, all cases show similar rate distributions. 
The rate is equal to the input value when the baseline is small,
but declines with increasing baselines. In particular, when the 
baseline is smaller than 20 days, the rates for both SN 1991bg-like
objects and SN 1991T-like objects equal the input values, regardless of
the extinction adopted for SN 1991T-like objects.

\section{Discussion}

{Throughout this paper we have categorized SN 1991T-like and 
SN 1991bg-like objects as ``peculiar" SNe Ia,  
for historical reasons: 
those objects were considered rare until a few years ago, and they show
distinct spectral features. There are discussions of
models for these ``peculiar" SNe Ia and whether they represent
physically distinct classes of SNe Ia having different explosion
mechanism and/or progenitor mass and configuration (e.g., Filippenko et 
al. 1992a, 1992b; Phillips et al. 1992; Ruiz-Lapuente et al. 1992; Jeffery
et al. 1992; Leibundgut et al. 1993;  Mazzali et al. 1995;
Turatto et al. 1996; Mazzali et al. 1997), but there is no 
concensus on the issue yet. It is also still an open question whether
SN 1991T-like and SN 1991bg-like objects are part of a spectroscopic
and/or photometric sequence, or constitute distinct observational 
classes. We thus emphasize that we subclassify SNe Ia as 
SN 1991T-like, normal, or SN 1991bg-like based only upon their
{\it apparent  spectroscopic abnormality}; we  do not imply
that they represent fundamentally different (but related) kinds of 
objects. 

It is also worth pointing out that after finding a 
high percentage (36\%) of 
SN 1991T-like and SN 1991bg-like SNe Ia (Li et al. 2000c),
it becomes questionable
whether to categorize these objects as ``peculiar." The classification
(normal or peculiar) 
of SN 1999aa-like objects, which comprised a significant 
portion of the SN 1991T-like objects in the sample of
Li et al. (2000c), also plays an important role. 
More detailed studies,  both observational and theoretical,
are needed to address these issues.

In our simulations we have employed a single light-curve shape
for each of the three subclasses of SNe Ia. One implication 
of this assumption is that SN 1991T-like objects always
have slower light curves than normal objects, which
conflicts with the observations of SN 1992bc (Maza et al. 
1994), a normal SN Ia having
a slow light curve ($\Delta m_{15}(B)=0.87$).  Hamuy (2000, private communication) 
also notes that SN 1999ee is a normal SN Ia with a slow
light curve. While the frequency of this
kind of normal SN Ia remains to be determined, it seems that the correlation
between light-curve shape and spectroscopic behavior is not
monotonic; two or more parameters may be needed to describe
the whole class of SNe Ia. Note, however, that the luminosity 
difference we adopted for the SN 1991T-like and the normal
objects (0.4 mag) is only slightly larger than the assumed 1$\sigma$ 
scatter (0.35 mag), which means that 
some objects overlap in luminosity between the two subclasses.
If we had adopted the luminosity vs. light-curve shape 
relation in our simulations, there would have been some normal SNe Ia
having large luminosity and thus having slow light curves.
Not doing this 
is therefore an apparent limitation of our simulations.

In the subsequent sections, we discuss the results of our
simulations in the $B$ band, more studies of the age 
bias, and comparisons of our results with observations.

} 

\subsection{The Results of the Simulations in the $B$ Band}

Although most modern surveys of nearby SNe
are done in unfiltered mode, for which our 
simulations using the $R$ band should represent the closest
match, some surveys are done in other passbands.
For example, CTSS was done in the $B$ band with photographic plates, and 
the SN surveys at high redshift are also done at the rest-frame $B$ band. 
To investigate the effect of adopting different passbands, we have also done
our simulations of magnitude-limited SN surveys in the $B$ band as both
CTSS and the high-redshift SN surveys are magnitude-limited (see Sec. 1).

The two major differences of the input to the Monte Carlo simulations 
of using the $B$ band are as follows. (1) The light 
curves of the SNe in the $B$ band are different. We have used the $B$-band 
light curves shown in Figure 12, where we adopted the data
of SN 1991T and SN 1991bg from Hamuy et al. (1996a), and the data
of the normal SN 1992al (Hamuy et al. 1996a). The rising parts
of the light curves are again constructed according to the recipe described
by Riess et al. (1999b), where we adopted risetimes of 22.0, 19.5, and
16.5 days for SN 1991T-like, normal, and SN 1991bg-like objects, respectively. 
(2) The differences among the luminosities of SNe Ia are different. In the 
$B$ band, SN 1991T-like objects are 0.4 mag brighter, and SN 1991bg-like
objects are 2.0 mag fainter, than normal SNe Ia. The difference of the 
adopted luminosities changes the input number of SNe to the simulations. 

The results of the simulations in the $B$ band are similar, but with 
subtle differences, to those 
in the $R$ band in terms of the apparent magnitude distribution (Figure 3),
the discovery efficiency at different baselines (Figure 4), and the
distribution of the epoch of discovery (Figure 5). The major difference,
however, is in the rates as shown in Figure 13. The SN surveys in the
$B$ band underestimate the rate of SN 1991bg-like objects more significantly
than do those in the $R$ band.

The rate of SN 1991T-like objects shows
some interesting differences. Since we have adopted equal luminosities for
the SN 1991T-like and the normal SNe Ia in the $R$ and $B$ bands, these 
differences in the observed rates are mostly caused by the different LCS
and age biases. When the baseline is small ($\lesssim$ 10 days), 
both the age bias and the LCS bias are small, so the surveys in 
the two passbands have the same peculiarity rate for the SN 1991T-like 
objects. When the baseline is moderately large (10 days $<$ baseline $\le$ 
33 days), the surveys in the $R$ band have larger rates for the SN 1991T-like
objects. The main reason for this is because at such 
baselines, the broader light curves of SN 1991T-like objects in the $R$ band
result in a relatively higher discovery efficiency for them, and yet the age
bias is not very severe at these baselines. When the baseline is very 
large ($>$ 33 days), the surveys in the $B$ band have higher rates
for the SN 1991T-like objects because of the more severe age bias
in the SN surveys in the $R$ band. This results from the fact that 
the $R$-band light curve is broader than the $B$-band light curve, so
a higher percentage of SNe Ia are discovered later than a week past maximum
in the $R$-band surveys.

\subsection{More About the Age Bias}

We have used a cutoff of 7 days past peak brightness 
to simulate the age bias. {For such a cutoff
we noticed that the age bias has negligible effects 
in magnitude-limited SN surveys (Figure 6) and distance-limited
SN surveys (Figure 11) when the baseline of the surveys is 
less  than 10 days, especially compared to the Malmquist and 
the LCS biases. This is reasonable since with such small 
baselines, very few SN 1991T-like objects are discovered
later than a week past maximum and classified as normal.}

We have also done simulations with 
different cutoffs (5, 3, 1, $-$1 days 
after peak brightness) to study the age bias,
but the results are not shown in Figures 3 through 6 and 
8 through 11 because of the increased complexity, 
making them too difficult to read.
Using different cutoffs for the age bias significantly 
affects the results of the simulations because of the changes in 
the number of SN 1991T-like objects classified by the observer as 
normal. The results for the SN 1991bg-like objects are not 
affected because they are not related to the SN 1991T-like
objects. The discovery efficiency of 
both the SN 1991T-like and the normal objects is affected
and the number distribution of SNe versus the 
epoch of discovery changes according to the adopted cutoffs.
The most important changes, however, are those for the 
rates of SN 1991T-like objects. 

Figure 14 shows the rates of SN 1991T-like objects
with different cutoffs for the age bias
for the magnitude-limited surveys simulated in the $B$ band. 
The rates of SN 1991T-like objects change dramatically with
the different adopted cutoffs -- the earlier the cutoff, the
smaller the rate of SN 1991T-like objects. For example, 
for the case SN 1991T ($A$=0.4) with a baseline of 20 days,
the rate of SN 1991T-like objects reduces from 16\% to 11\% to
 6\%, when the cutoff changes from 7 to 3 to $-$1 days 
after peak brightness. The rate of SN 1991T-like objects
is only 4\% for the case SN 1991T ($A$=0.8) when the adopted
cutoff for the age bias is $-$1 day after (i.e., 1 day
before) peak brightness.

Figure 15 shows similar results for the distance-limited surveys 
simulated in the $R$ band. We notice that when the limiting 
magnitude is deep enough and the baseline is small enough 
(e.g., baseline $\le$ 10 days for the limiting magnitude 
19.0 cases, and  baseline $\le$ 5 days for the limiting
magnitude 17.0 cases {except the case with $A$=0.8}),   the
rate of SN 1991T-like objects is still equal to the intrinsic
value, regardless of the adopted (reasonable) cutoff and 
extinction. The reason for this is that in those cases, all SN
1991T-like objects are discovered prior to 1 day before peak
brightness, so the observed luminosity function is not 
affected by the selected cutoff for the age bias or
by the adopted extinction. When the baseline becomes larger, however,
the earlier the adopted cutoff for the age bias,
the smaller the rate of SN 1991T-like objects. When the limiting
magnitude of the survey is bright (e.g., the mag 15.0 cases in Figure 15), 
the survey is actually magnitude-limited as discussed earlier,
and the results reflect this for the magnitude-limited surveys in
the $R$ band. Comparison with Figure 14 reveals that they are very
similar. 

As the cutoff of the age bias significantly affects
the results of the simulations, we have tried to
constrain it from observations. Unfortunately,
there are no spectroscopic observations of SN 1991T 
in the literature between $-$3 and +9 days (e.g., Filippenko
et al. 1992a; Phillips et al. 1992a; Ruiz-Lapuente et al. 1992), nor
are there any observations of SN 1997br, another 
SN 1991T-like object (Li et al. 1999), between $-$4 and +8 days. 
We can see that the spectra
of the two objects are very different from those of 
normal SNe Ia at $-$3 or $-$4 days, so the cutoff must 
be later than $-$3 days. Their spectra are relatively similar to 
those of normal SNe Ia after 8 days, so the cutoff 
probably should be earlier than 8 days. {(Notice that
SNe 1991T and 1997br are slightly different from normal SNe 
Ia at late times, as discussed by Filippenko et al. [1992a]
and Li et al. [1999], but these
differences are subtle and the spectra do not show the defining 
features of SN 1991T-like objects -- strong Fe III lines and 
no/weak Si II 6150 \AA).}
This range for the cutoff ($-2$ to 7 days)
is large and does not provide much  of a 
constraint for useful discussions. {Li et al. (2000c)
compared the spectra of three SN 1999aa-like objects near 
maximum to those of normal SNe Ia, and suggested that
the cutoff of the age bias 
might be as early as the day of peak brightness or perhaps 
1 day earlier. This suggestion, however, depends on 
whether SN 1999aa-like objects should be considered in the 
SN 1991T subclass or as ``normal" SNe Ia. Further studies of
SN 1999aa-like objects and their implications for spectral 
classification of SNe Ia are needed (e.g., Li 
et al. 2000d).}

{As the database of SN Ia spectra becomes better, peculiar 
ones will become easier to identify. For example, we may not even need 
the defining characteristics (Fe III lines and no/weak 
Si II 6150 \AA) to identify a SN 1991T-like 
object. Thus, the age bias will decrease with the observer's
increasing ability to identify peculiarities in the spectrum. }

\subsection{Comparisons with Observations}

Our simulations in the $R$ band indicate that a distance-limited survey with a deep
limiting magnitude and a small baseline, such as LOSS and BAOSS
(Li et al. 2000c), will discover all the SNe Ia
in the sample without any bias against any peculiar types of SNe.
The uncertain determination of quantities such as the extra extinction
of SN 1991T-like objects and the cutoff of the age bias
also plays a  negligible role.
Li et al. derive the intrinsic peculiarity rate and 
luminosity function of SNe Ia from 45 SNe Ia discovered in
the course of LOSS and BAOSS. They find a high rate for peculiar
SNe Ia (about 36\%), and that the luminosity function of 
SNe Ia has 20\% SN 1991T-like, 64\% normal, and 
16\% SN 1991bg-like objects.

It is also interesting to compare the results of our simulations 
with those of real SN surveys. Figure 16 compares
the distribution of the epoch of discovery from our simulations
with those from LOSS$+$BAOSS and from CTSS. 
As indicated by Li et al. (2000c), LOSS and BAOSS are done in the
unfiltered mode and are distance-limited
surveys with a limiting magnitude of 19.0 and a baseline of $\le$ 10 days, 
and the limiting distance of the sample galaxies is very similar
to what we used in our simulations (where normal SNe Ia have a peak
apparent magnitude of 16.0). CTSS is done 
in the $B$ band and is a magnitude-limited
survey with a baseline of about 20 days (Hamuy \& Pinto 1999). Figure 
16 shows that the results of the real SN surveys are quite consistent
with those of our simulations.

Hamuy \& Pinto (1999) have used the same approach
(that is, Monte Carlo simulations) to study 
the observational biases in CTSS. They simulated
the survey as  magnitude-limited, and took into consideration 
the LCS bias and the Malmquist bias. They also considered the 
Shaw (1979) effect, which requires the discovered SN to have
a minimum projected separation from the nucleus of the host galaxy
and is an important observational bias in photographic 
searches such as CTSS. They used the real
observation history of the survey to do the simulation, and a
luminosity function of SNe Ia based on the decline rate [that is, 
$L$ vs. $\Delta m_{15}(B)$].
The age bias, however, was not considered in their simulations. 
They estimated the degree of incompleteness of the survey as a function
of decline rate, and corrected the luminosity function of SNe Ia.
They found that the true luminosity function of SNe Ia is remarkably
flat over the whole range of decline rates, with perhaps an increase
in frequency for the intrinsically faint events. 

Adopting the luminosity function of SNe Ia as derived by Li et al.
(2000c) and applying the results from our simulations, we find that 
for a magnitude-limited SN survey in the $B$ band  with a baseline of 20 days, SN 1991bg-like
and SN 1991T-like objects should comprise about 1\% and 27\% of 
the total sample, respectively (assuming SN 1991T-like objects do not 
suffer more extinction than the other SNe Ia, and adopting
a cutoff of 7 days after
peak brightness for the age bias). This means that for
the 31 SNe discovered
in CTSS, there should be 0.3$\pm$0.5 SN 1991bg-like
objects  and 8.4$\pm$2.9 SN 1991T-like objects (the 
uncertainties are derived from Poisson statistics).
It is natural that
there is one SN 1991bg-like object in the Cal\'an/Tololo  sample 
(SN 1992K; Hamuy et al. 1994), but the fact that no objects were classified
as SN 1991T-like is somewhat surprising.

There are several possible reasons
for the apparent lack of SN 1991T-like objects. (1) As discussed 
earlier, they seem more likely to occur in dusty, 
star-forming regions than the other two kinds of SNe Ia. This causes two
difficulties for discovering SN 1991T-like objects in CTSS.
First, the star-forming
regions are more likely to be overexposed in the photographic plates, 
resulting in an effect similar to that of Shaw (1979).
Second, the dusty
environment of SN 1991T-like objects causes them to be more 
extinguished on average.
(2) As also previously discussed, a more serious age 
bias (e.g., a cutoff of 1 day before peak brightness) significantly
reduces the apparent rate of SN 1991T-like objects. The rate of SN 1991T-like 
objects changes to only 6\% and 4\% for the 
cases SN 1991T ($A$=0.4) and  SN 1991T ($A$=0.8),
which means there should be 1.9$\pm$1.4 and 1.2$\pm$1.1 SN 1991T-like objects
in CTSS, respectively.
The statistical significance of the null discovery of SN 1991T-like 
objects in CTSS thus depends on the still
uncertain cutoff of the age bias, but we believe this is the main bias
affecting CTSS. 

\section{Conclusions}

The differences among the luminosities, light-curve shapes, and spectroscopic
evolution of the three major subtypes of SNe Ia (SN 1991T-like, normal, and 
SN 1991bg-like) result in several observational biases in SN 
surveys: the age bias, the Malmquist bias, the light-curve
shape bias, and the extinction bias. These biases affect the observed peculiarity 
rate and luminosity function of SNe Ia. 
Under a number of assumptions, we have conducted Monte Carlo simulations of the observations of SNe Ia in the $R$ band in 
magnitude-limited and distance-limited SN surveys to study the effect of 
the biases and to investigate the properties of the surveys
having different parameters.

We have simulated magnitude-limited SN surveys with a limiting magnitude of 19.0
and a baseline of 1 to 60 days. We find that with our assumptions, the 
SN discovery efficiency can only reach a maximum of about 84\%, and the longer the
baseline, the more the SNe are lost and the more serious is the age bias. 
We also find that the majority of SNe Ia could be detected before maximum brightness
only when the baseline of the surveys is $\lesssim$ 10 days. We find that 
the magnitude-limited surveys significantly underestimate the 
rate of SN 1991bg-like objects, and that the adopted extra
extinction for SN 1991T-like objects plays an important role in determining
their rate. The surveys overestimate the rate of SN 1991T-like objects
if no extra extinction is assumed for them, and underestimate their 
rate if the adopted extra extinction is larger than 0.4 mag. 

We have also simulated distance-limited surveys with a limiting distance
where normal SNe Ia have a peak apparent magnitude of 16.0. Results for 
surveys with limiting magnitude from 15.0 to 19.0 and baseline from 1 to 60
days have been obtained. We find that the discovery efficiency increases with
increasing limiting magnitude, and when the limiting magnitude is faint enough
(e.g., 19.0) and the baseline is short enough (e.g., $\lesssim$ 10 days), all
SNe Ia are discovered in a survey.  This means that the 
observed peculiarity rate and luminosity function of SNe Ia in such
surveys will be the intrinsic ones.  The roles of the LCS bias
and the age bias become more important when the baseline 
is longer and the limiting magnitude is brighter. 

{It is worth mentioning that although magnitude-limited surveys
can bias significantly the luminosity function of SNe Ia,  a carefully
selected subsample of SNe Ia found in a magnitude-limited
survey can actually obey the rules of a distance-limited survey, 
and thus can produce an unbiased luminosity function of SNe Ia
if the baseline of the survey is small enough. }

We have done the simulations in the $B$ band and 
find that these SN surveys underestimate the SN 1991bg-like objects
more significantly than do those in the $R$ band. The LCS bias and 
the age bias for the SN 1991T-like objects are also different
in the two passbands, resulting in a different dependence of their peculiarity
rate on baselines. We find that the rate of SN 1991T-like objects in 
the simulations is significantly affected by the adopted cutoff for
the age bias. The distance-limited SN surveys with 
deep limiting magnitude and short baseline remain 
the best ones for determining the intrinsic peculiarity
rate and luminosity function of SNe Ia, as these quantities are
not affected by the poorly constrained  extra extinction of the 
SN 1991T-like objects or by the adopted cutoff of the age bias.

We have compared the distribution of the epoch of discovery from our
simulations with those of real SN surveys, specifically LOSS$+$BAOSS and 
CTSS, and find that they are quite consistent. 
We have also compared our simulations to those done by Hamuy \& Pinto (1999). 
The lack of SN 1991T-like objects in CTSS 
is somewhat mysterious, and may be caused by 
the Shaw effect, extra extinction to such objects, and perhaps
most importantly the age bias.

\acknowledgements{The authors thank the referee, M. Hamuy, for his 
very helpful comments on the paper. We also thank P. Pinto and
M. M. Phillips for many
useful discussions.  This work is supported by NSF 
grants AST-9417213 and AST-9987438, as well as 
by NASA grant AR-08006 from the Space
Telecope Science Institute, which is operated by AURA, Inc.,
under NASA contract NAS5-26555. We are also grateful to the Sylvia
and Jim Katzman Foundation for generous donations that made LOSS
and the resulting analysis possible.
}

\newpage

\begin{figure}
\caption{ The $R$-band light curves of SN 1991T-like, normal, and 
SN 1991bg-like objects used in the Monte Carlo simulations.}
\label{1}
\end{figure}

\begin{figure}
\caption{The input number distribution of SNe Ia versus apparent
magnitude at maximum brightness in the magnitude-limited survey. SNe are binned to an interval of 0.01 mag.}
\label{2}
\end{figure}

\begin{figure}
\caption{The number distribution of discovered SNe Ia versus 
apparent magnitude at maximum brightness in the magnitude-limited surveys. Only two cases are shown here: 
SN 1991bg-like objects, and SN 1991T ($A$=0). 
The upper panel shows the distribution of the peculiar SNe Ia,
while the lower panel shows the corresponding distribution of
the normal objects. Results for different baselines are shown
for each type of SN. The dotted lines labeled ``perfect" mark the
expected number distributions if all SNe are discovered. }
\label{3}
\end{figure}

\begin{figure}
\caption{The discovery efficiency for the magnitude-limited surveys found in the simulations. The upper
panel shows the discovery efficiency for the SN 1991bg-like and the normal
objects, the middle panel shows it for the SN 1991T-like 
and the normal objects, while the lower panel shows it for the 
SN 1991T-like and the normal objects without considering the age
bias. The dashed line is the discovery efficiency 
for the normal objects for the case SN 1991T ($A$=0.4), and the dotted
line for the case SN 1991T ($A$=0.8). 
}
\label{4}
\end{figure}

\begin{figure}
\caption{ The number distribution of SNe versus the epoch of discovery for the
magnitude-limited surveys. (Epoch = 0 is defined to be the day of 
 $B$ maximum brightness.)
Only two cases are illustrated here: SN 1991bg-like objects, 
and SN 1991T ($A$=0). 
The upper panel shows the distribution of the peculiar SNe Ia, while
the lower panel shows the corresponding distribution of the 
normal objects. Results for different baselines are given. }
\label{5}
\end{figure}

\begin{figure}
\caption{The peculiarity rate in the magnitude-limited surveys found in the simulations. The upper
panel shows the result for the SN 1991bg-like objects, while the lower
panel shows results for the SN 1991T-like objects. The dash-dotted line
is the input rate for each type of supernova.}
\label{6}
\end{figure}

\begin{figure}
\caption{The input number distribution of SNe Ia versus apparent
magnitude at maximum brightness in the distance-limited surveys.
SNe are binned to an interval of 0.01 mag. The dashed line is
the number distribution for the case SN 1991T ($A$=0.4) and the 
dotted line is that for the case SN 1991T ($A$=0.8).
}
\label{7}
\end{figure}

\begin{figure}
\caption{The number distribution of discovered SNe Ia versus 
apparent magnitude at maximum  for the distance-limited surveys. Only the case
of SN 1991T ($A$=0.8) is shown here for different limiting magnitudes with 
different baselines. The dotted lines mark the expected
number distribution if all SNe are discovered. In the lower two 
panels they are indistinguishable from the results of 
the simulations. }
\label{8}
\end{figure}

\begin{figure}
\caption{The discovery efficiency for the distance-limited surveys. The 
upper panel shows the discovery efficiency for the peculiar SNe Ia, while
the lower panel shows the corresponding efficiency for the normal ones. }
\label{9}
\end{figure}

\begin{figure}
\caption{The number distribution of SNe versus the epoch of discovery for the
distance-limited surveys. (Epoch = 0 is defined to be the day of $B$ maximum 
brightness.)
Only two cases are shown here: SN 1991bg-like objects, 
and SN 1991T ($A$=0.8). For each case, results for different
baselines are illustrated with different limiting magnitudes.}
\label{10}
\end{figure}

\begin{figure}
\caption{The peculiarity rates in the distance-limited surveys
found in the simulations. Results are shown for different baselines
and different limiting magnitudes. The solid, dashed, and dash-dotted 
lines are for the 
cases of a limiting magnitude of 19.0, 17.0, and 15.0, respectively. 
The dotted lines are the input rate.}
\label{11}
\end{figure}

\begin{figure}
\caption{ The $B$-band light curves of SN 1991T-like, normal, and 
SN 1991bg-like objects used in the Monte Carlo simulations.}
\label{12}
\end{figure}

\begin{figure}
\caption{ Comparison of the results of the peculiarity rate in the 
magnitude-limited surveys in the $B$ and $R$ passbands. The upper panel
shows the result for the SN 1991bg-like objects, while the lower panel 
gives results for the SN 1991T-like objects. The dash-dotted line is 
the input rate for each type of SN. The lines with solid circles are
those for the results in the $B$ band, while the others are for those
in the $R$ band. }
\label{13}
\end{figure}

\begin{figure}
\caption{The rates of SN 1991T-like objects in the magnitude-limited
surveys with different cutoffs for the age bias. In each panel
the solid, dashed, and dash-dotted  lines are
the results for cutoffs of 7, 3, and $-$1 days past maximum brightness,
respectively. The dotted lines are the input rate.}
\label{14}
\end{figure}

\begin{figure}
\caption{The rates of SN 1991T-like objects in the distance-limited
SN surveys with different cutoffs for the age bias. From left
to right: The limiting magnitude of 19.0, 17.0, and 15.0 cases;
from top to bottom: the cases SN 1991T ($A$=0), ($A$=0.4), 
($A$=0.8). In each panel the solid, dashed, and dash-dotted lines are
the results for 
a cutoff of 7, 3, and $-$1 day past maximum brightness for
the age bias, respectively. The dotted lines are the input rate.}
\label{15}
\end{figure}

\begin{figure}
\caption{Comparison of the distribution of the epoch of discovery from 
the simulations with that from LOSS$+$BAOSS and CTSS. The upper panel shows the results 
from CTSS ({\it solid line}) and a magnitude-limited
survey with a baseline of 20 days ({\it dashed line}), while the lower panel shows 
those from LOSS$+$BAOSS ({\it solid line}) and a distance-limited survey
with a baseline of 10 days and a limiting magnitude of 19.0.
}
\label{16}
\end{figure}

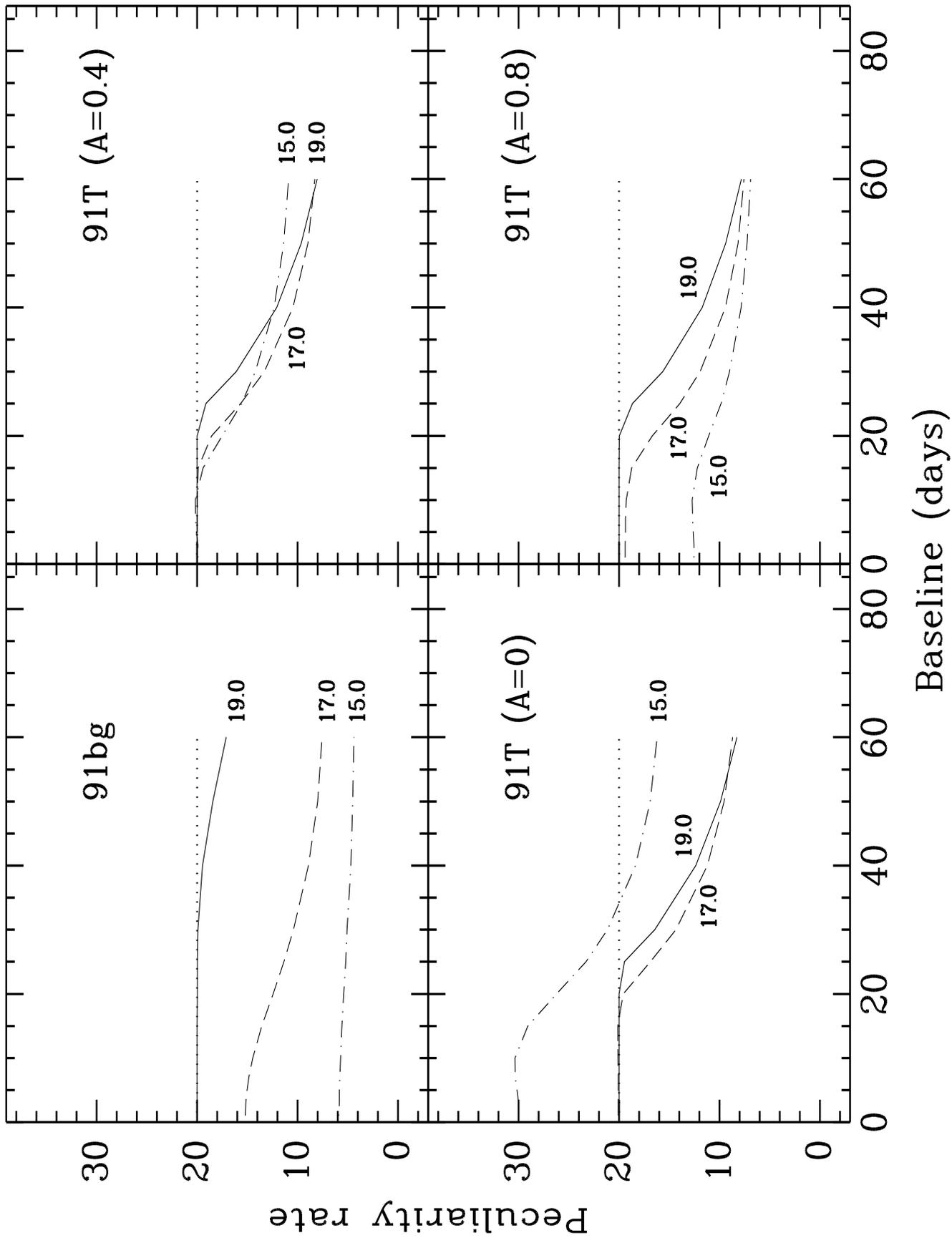
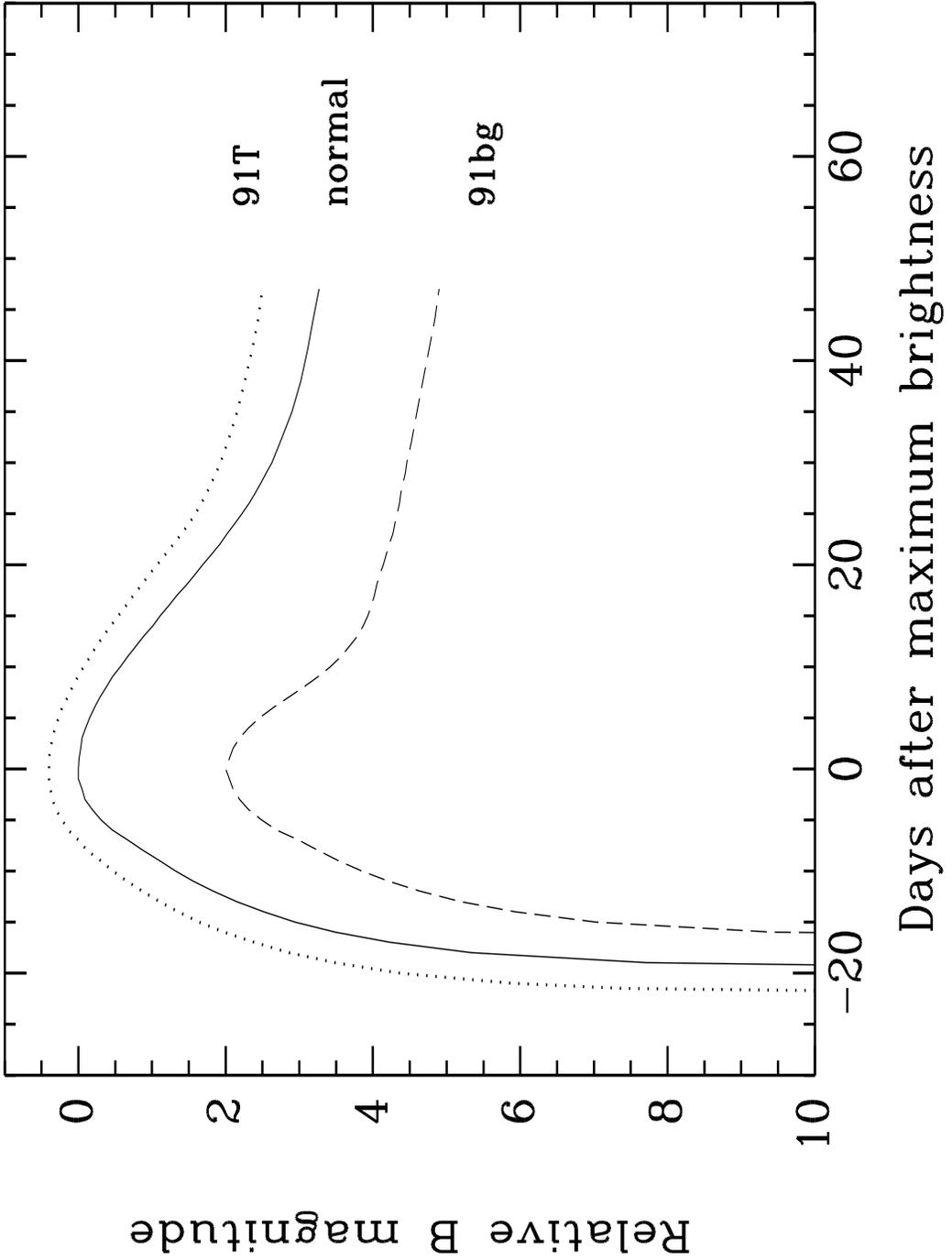
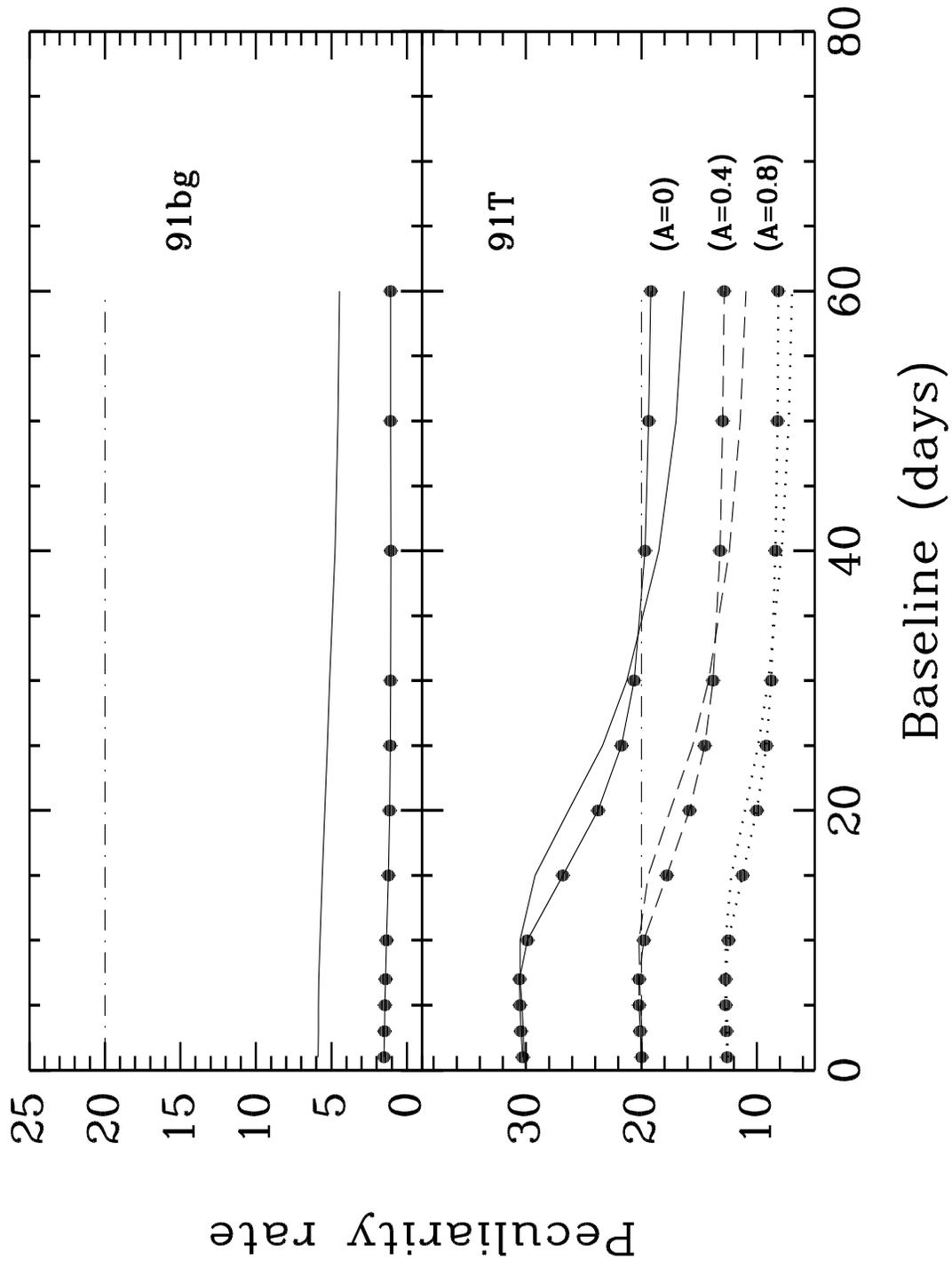
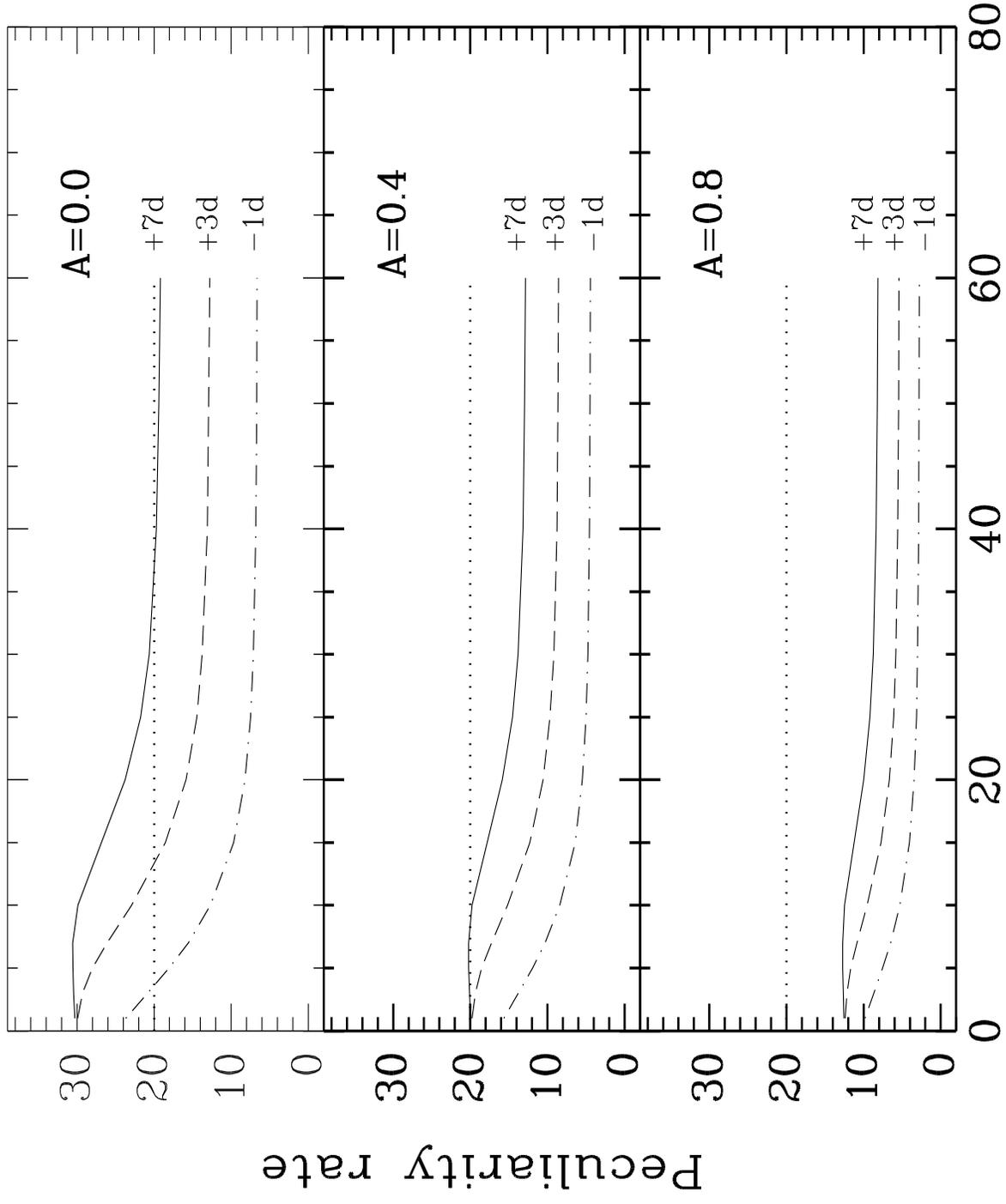
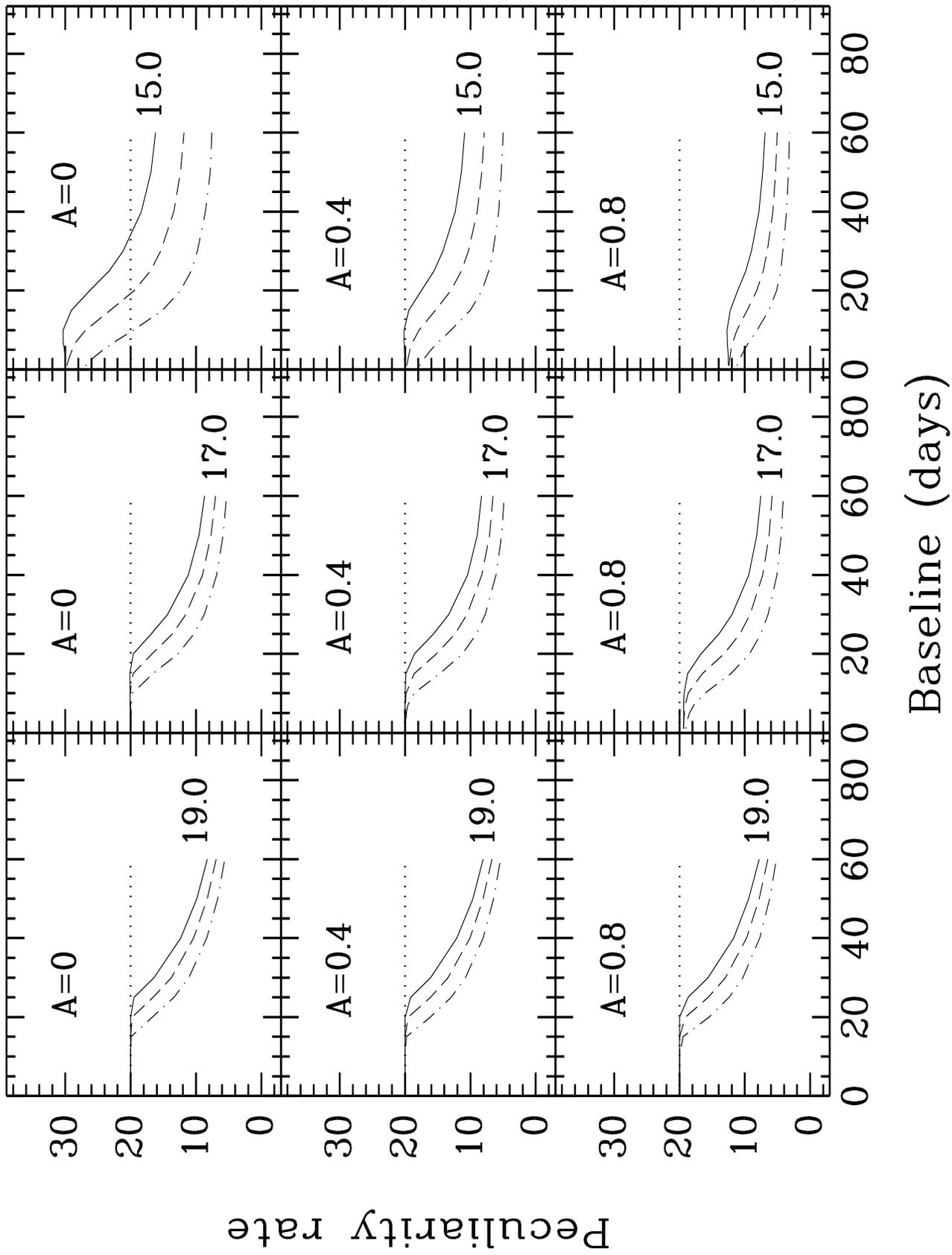
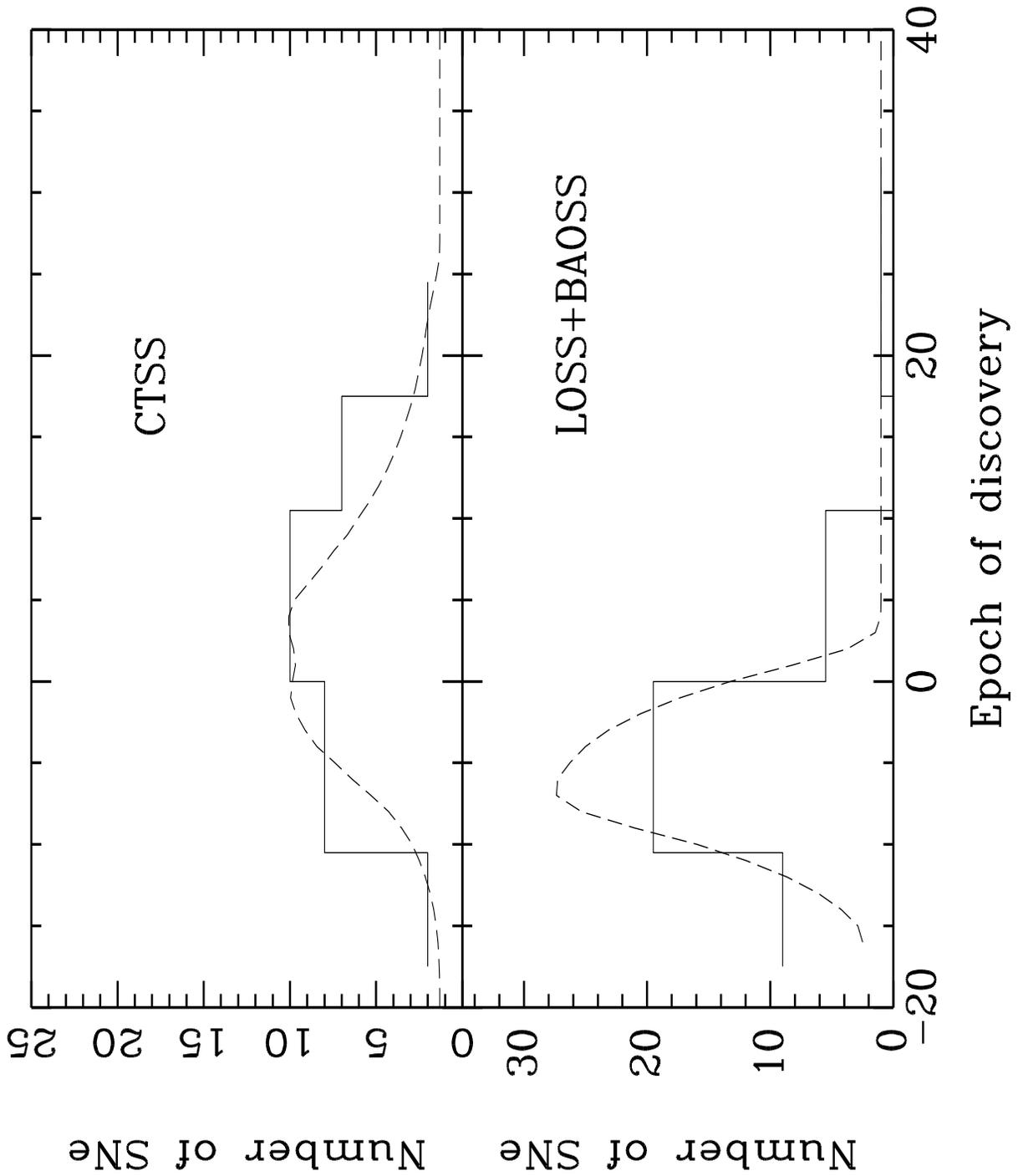
\end{document}